\documentclass{iopart}

\usepackage{amssymb}
\usepackage{url}
\usepackage{citesort}

\newcommand{\eqref}[1]{{(\ref{#1})}}

\renewcommand{\paragraph}[1]{{\vspace{3pt} \noindent \textit{#1}}}
 
\begin{document}

\title{A LISA Data-Analysis Primer}

\author{Michele Vallisneri}

\address{Jet Propulsion Laboratory, California Institute of Technology, Pasadena, CA 91109, USA}

\ead{Michele.Vallisneri@jpl.nasa.gov}

\begin{abstract}
This article is an introduction for the nonpractitioner to the ideas and issues of LISA data analysis, as reflected in the explorations and experiments of the participants in the Mock LISA Data Challenges. In particular, I discuss the methods and codes that have been developed for the detection and parameter estimation of supermassive black-hole binaries, extreme mass-ratio inspirals, and Galactic binaries.
\end{abstract}

\vspace{-0.75cm}


\section{Introduction}


The Laser Interferometer Space Antenna (LISA) is a joint NASA--ESA mission to measure gravitational waves (GWs) in the low-frequency band between a few $10^{-5}$ and $10^{-2}$ Hz. LISA consists of a triangular constellation of spacecraft, flying in a solar orbit, and exchanging laser beams to monitor their relative distances ($\sim 5 \times 10^6$ km) to picometer precision \cite{lisa}.  The LISA dataset will contain GWs from millions of sources, most of them in our Galaxy, but many in the distant Universe, as far as $z \sim 20$.  Some sources, such as Galactic binaries, have simple waveforms; some, such as the inspirals of supermassive black holes (SMBHs), will be so strong that they will stand out clearly in the dataset. Other sources, such as the extreme--mass-ratio inspirals (EMRIs) of compact objects into central galactic BHs, generate exceedingly rich, complicated, but relatively weak waves that we will need to dig out from the LISA noise by matching their phases accurately across $\sim 10^5$ cycles. The millions of compact binaries in our Galaxy will merge in a confusion-limited foreground that we will be able to characterize statistically, and from which we will individually resolve a fraction of louder or higher-frequency sources.

It is clear from this picture that effective data-analysis techniques will be crucial to the success of this mission, which hopes to deliver an impressive list of scientific payoffs, ranging from characterizing the hierarchical growth of galaxies and their central BHs across the history of the Universe, to testing the strong-gravity limit of general relativity in MBH mergers; from mapping the Kerr geometry of spacetime around MBHs, to surveying the demographics of compact binaries in our Galaxy; and much more. (The LISA science-case document \cite{lisascience} is a very good place to start learning and to find extensive references about the science that will be enabled by LISA.) In fact, LISA data analysis should be construed as an integral part of its measurement concept, since LISA's true science data will be the counts and parameters of the GW sources it detects.

The Mock LISA Data Challenges (MLDCs)
\cite{mldc1form,2007CQGra..24..529A,2007CQGra..24..551A,2008CQGra..25k4037B,2008CQGra..25r4026B},
begun in 2006, are a community-wide effort to demonstrate and advance LISA data-analysis capabilities, understanding its conceptual and quantitative peculiarities, and kickstarting the development of algorithms and software.  In each round of the MLDCs, \emph{challenge} datasets are released that contain GW signals from sources of undisclosed parameters, embedded in synthetic LISA noise; challenge participants have a few months to analyze the data and submit detection candidates, which are then compared with the sources originally injected in the datasets. (\emph{Training} datasets with public source parameters are also provided, to help participants tune and troubleshoot their searches.) 

At the time of this symposium, three MLDC rounds have been completed: MLDC 1 \cite{mldc1form,2007CQGra..24..529A} focused on establishing basic techniques to observe GWs from compact Galactic binaries, isolated or moderately interfering; as well as from the inspiral phase of bright and isolated nonspinning MBH binaries.  The more ambitious MLDC 2 \cite{2007CQGra..24..551A,2008CQGra..25k4037B} featured a dataset containing GW signals from approximately 26 million Galatic binaries generated with a population-synthesis code; a dataset (the ``whole enchilada'') with a similar population, plus GW signals from a small number of nonspinning MBH binaries and EMRIs; and five more datasets with single-EMRI signals.  MLDC 1B \cite{2008CQGra..25r4026B} was a repeat of MLDC 1 (with the addition of single-EMRI datasets), meant as a second entry point to the MLDCs for new, less experienced participants.  Last, MLDC 3 \cite{2008CQGra..25r4026B}, now in progress, marks the inclusion of more realistic source models (such as chirping Galactic binaries and spinning MBH binaries) and of new types of sources  (such as short-lived bursts and stochastic backgrounds).

This article is a brief introduction to the ideas and techniques of LISA data analysis, as exemplified by the published reports authored by MLDC participants about their work. Collectively, those reports illuminate a broad swath of the state of the art in this field. More informal discussions are also available in the form of the technical notes submitted with MLDC entries \cite{mldcreports}. I shall focus on the methods, rather than their quantitative results; for those, see \cite{mldcreports,2007CQGra..24..529A,2008CQGra..25k4037B,2008CQGra..25r4026B}, and the MLDC articles cited below. Furthermore, I shall concentrate on the analysis of LISA data, as opposed to the equally crucial modeling of LISA sources. The approach taken in the MLDCs is that the development of data-analysis algorithms can proceed in parallel to source-modeling research, by assuming that currently available waveforms and source models can be used as stand-ins for more accurate versions that will be available by the time LISA returns data.

In the rest of this paper, I briefly describe the peculiarities of LISA data analysis compared to ground-based observatories (section \ref{sec:lisadata}), and I discuss the plans to detect and characterize the three class of sources that have so far been the focus of the MLDCs: SMBHs (section \ref{sec:smbh}), EMRIs (section \ref{sec:emri}), and Galactic binaries (section \ref{sec:gbs}). Thus, I do not cover the recent progress toward other LISA data-analysis objectives, such as detecting bursts from cosmic strings \cite{keycornish2008} and characterizing stochastic backgrounds (see \cite{stochasticprinciple} for the basic detection principles, \cite{2008CQGra..25r4019R} for a simple implementation, and \cite{stochasticanisotropy} for the prospects of constraining anisotropy and polarization).


\section{The LISA dataset}
\label{sec:lisadata}

A good way to begin to understand the peculiarities of LISA data analysis is to contrast the LISA dataset with those of ground-based experiments such as LIGO \cite{1999PhT....52j..44B}.

\paragraph{The dataset is dominated by GW signals,} whereas ground-based detector data are dominated by instrument noise. Thus individual LISA sources must be detected over the background of many others, which raises issues of confusion between similar signals, and complicates the attribution of statistical significance to detections.

\paragraph{The LISA dataset is small---$5 \times 10^8$ samples} ($1$ Hz bandwidth $\times$ 5 yr operation $\times$ 3 independent GW observables), to be compared with $\sim 4 \times 10^{11}$ samples for the one-year LIGO S5 run alone. The difference originates mainly in the $10^3 : 1$ ratio of bandwidths, and it points to a different character of data analysis: a fine-tuning job of fitting many source parameters to relatively few bits for LISA, rather than a needle-in-the-haystack data-mining problem for ground-based detectors.

\paragraph{The LISA response is complex, and it is key to LISA's directional sensitivity.}
For both LISA and ground-based detectors, incoming GW signals are periodically Doppler shifted by the motion of the detector around the Sun, and modulated by the rotation of its sensitive axes. Because most LISA sources are long lasting, these effects imprint GWs with information about the sky location and orientation of sources. By contrast, sources for ground-based detectors are short lived (with the exception of pulsars), so sensitivity to source position is provided by the triangulation of times of arrival in a detector network.\footnote{Triangulation is relevant to LISA in the higher-frequency part of its sensitive band, where GW wavelength is comparable to the LISA armlengths, so the LISA output is complicated by significant phase delays between the response of the three spacecraft to the waves.} Thus, the signal models used in LIGO searches do not generally include sky position as a parameter, while the signal models for LISA must.

\paragraph{One LISA is equivalent to a network of three synthesized GW interferometers.} To cancel the otherwise overwhelming laser frequency noise, the phase measurements taken between LISA spacecraft must be carefully assembled into the observables of \emph{time-delay interferometry} (TDI; see \cite{2005PhRvD..72d2003V} for an introduction), which complicates the LISA response even more. Many different observables can be formed; however, when all six laser links are operational, LISA can be described as producing \emph{two} independent, GW-sensitive measurements for frequencies below $2 \times 10^{-3}$ Hz, and three above.\footnote{A simple way to think about this is that three LIGO-like interferometers can be formed with the three LISA spacecraft, but at long wavelengths the sum of the three signals must be zero by symmetry, which removes one independent measurement.}

\section{Supermassive--black-hole binaries}
\label{sec:smbh}

Observations of galactic nuclei from the ground and space suggest that SMBHs reside at the center of most galaxies. Hierarchical structure formation models explain SMBH growth as the result of the mergers of central BHs in merged galaxies \cite{2006ApJS..163...50H}; we observe many of the latter. LISA is expected to detect GWs from a few to a hundred SMBH inspirals per year, with masses between $10^4$ and a few $10^7 \, M_\odot$, and from redshifts as far as $z \sim 20$ \cite{2007MNRAS.377.1711S}.

\paragraph{Signal modeling.} Binary-coalescence waveforms have been customarily divided in the phases of inspiral, merger, and ringdown; rather than a true physical distinction, however, this division has echoed a specialization of effort in the GR community, and the use of different mathematical techniques to compute the corresponding pieces of waveforms (viz., post--Newtonian equations, the direct numerical integration of the Einstein equations, and perturbation methods on BH backgrounds).

As for data analysis, searches on data from ground-based experiments have so far targeted inspiral and ringdown waveforms (from binaries of BHs, as well as other stellar remnants; see, e.g.,
\cite{ligosearch,2008PhRvD..78d2002A},
because theoretical models for these partial waveforms have been easily available as analytical functions of the source parameters.  Numerical relativity's recent success in robustly computing several GW cycles from a variety of merger configurations (see
\cite{breakthrough}
for the breakthrough results, and \cite{pretoriusreview} for a review) has rekindled the interest in implementing inspiral--merger--ringdown searches for ground-based experiments. Given the substantial computational cost of merger waveforms, these searches rely on analytical waveform families obtained by fitting to ``stitched'' post--Newtonian and numerical waveforms, computed for a small number of parameter sets \cite{stitched}.
The collaboration between numerical relativists and GW data analysts has recently become tighter thanks to the Numerical INJection Analysis (NINJA) project \cite{ninjasite}, whereby numerical waveforms produced by ten different codes have been injected in synthetic instrument noise, and then searched for using the LIGO--Virgo Collaboration inspiral and burst pipelines, as well as more heterodox techniques such as the Hilbert--Huang transform.

The experience accreted in ground-based searches for BH binaries will transfer directly to LISA data analysis, not least because waveforms for binaries in the LISA frequency range, with total masses $M = 10^4\mbox{--}10^8 \, M_\odot$, are time-rescaled versions of the waveforms for stellar-mass binaries in the ground-based range (this is because $M$ is effectively the only timescale in the problem, while all other source parameters are or can be made dimensionless). However, additional work will certainly be needed on the source-modeling side, because waveform-accuracy requirements are much less stringent for ground-based detectors than for LISA: the former will yield detections at modest signal-to-noise ratios (SNR) that limit the resolving power of searches and therefore increase tolerable waveform errors; the latter will detect SMBH binaries with much higher SNRs, so accurate waveforms (including the effects of orbital eccentricity \cite{2006CQGra..23S.785B} and spin interactions and waveform corrections at higher post--Newtonian order \cite{2007PhRvD..75l4002A}) will be required to estimate source parameters reliably \cite{2007PhRvD..76j4018C} and to avoid biasing the detection of other weaker sources in the LISA data.

\paragraph{Matched filtering.} The possibility of computing accurate waveforms for BH binaries as functions of a small number of parameters, as well as the phase coherence of these signals across their entire duration, make them ideal targets for detection by matched filtering (see, e.g.,
\cite{basics}).
This is the gold standard of GW data analysis, in which a set of theoretical waveform models (\emph{templates}), computed for source-parameter values spread across their likely ranges, are individually subtracted from the measured detector output, and the statistical likelihood of the residuals is evaluated with respect to a model of detector noise, assumed additive. If the noise is Gaussian and stationary, the likelihood is a function of a single number, the \emph{SNR after filtering} $\rho(h_i) = (s,h_i) / \sqrt{(h_i,h_i)}$: here $s$ is the noisy signal, $h_i$ is a template, and the inner product 
\begin{equation}
(g_1,g_2) = 4 \, \mathrm{Re} \int_0^{+\infty} \frac{g_1^*(f) g_2(f)}{S_n(f)} \, \rmd f
\label{eq:innerproduct}
\end{equation}
is a generalization of a time-domain correlation product. Equation \eqref{eq:innerproduct} is weighted by the noise's one-sided power spectral density $S_n(f)$ toward the frequencies at which the detector is most sensitive. A detection can be claimed for template $h_i$ when $\rho(h_i)$ is larger than a threshold $\rho^*$ chosen to yield a negligible rate of false alarms. In fact, GW signals will ``ring'' several templates in a neighborhood of parameter space, so the detection is attributed to the template with the maximum SNR.

In the practice of ground-based searches, detector noise can be glitchy and nonstationary, so additional detection criteria are introduced: coincidence between different detectors, data-quality vetos that exclude noisy or problematic stretches of data, 
and signal-based vetos such as $\chi^2$, which compares the contribution of different frequency bands to $\rho$. Moreover, the statistical value of indicators such as $\rho$ is not assumed \emph{a priori} from noise models, but it is derived from a frequentist analysis of \emph{backgrounds} (how often do high-SNR candidates result from noise alone?) and \emph{detection efficiencies} (how often does the search pipeline recover GW signals artificially injected into the data?).
It is unclear to what extent such cautions will be needed in LISA searches: while they are certainly less important for high-SNR detections, little is known at this time about the Gaussianity and stationarity of the LISA noise, and a full understanding may be possible only at the time of commissioning. The availability of multiple TDI observables will certainly play a role in the eventual solution \cite{Bender:2006rz}. From the viewpoint of LISA data-analysis development, it is therefore reasonable to concentrate early efforts on Gaussian-noise searches, which in any case make up the basic framework and dictionary for more careful statistical studies.

\paragraph{Template placement.} The most important decision that must be made in implementing a matched-filtering search is how to distribute the templates to be tested across source-parameter space. Ground-based searches for GWs from nonspinning compact-object binaries have so far placed templates along structured grids (rectangular, or hexagonal) in the plane of the two binary masses $m_1$ and $m_2$. The grids are set densely enough that, for any true GW signal present in the data, the nearest template would still recover a large percentage (the \emph{minimum match}) of the \emph{optimal} SNR achieved by a perfectly matching template. In practice, the placement is performed by finding new coordinates $c_1(m_1,m_2)$ and $c_2(m_1,m_2)$ such that a regular grid in $c_1$ and $c_2$ achieves the minimum-match constraint approximately homogeneously across the mass plane.

Building such reassuringly exhaustive \emph{template banks} has been possible for signal models that depend only on two parameters---technically there are a few more, such as the strength, time of occurrence, and initial phase of the inspiral waves, but $\rho$ can be maximized analytically over these \emph{extrinsic} parameters without need for distinct templates to be placed along their ranges. For larger number of \emph{intrinsic} parameters, producing homogeneous banks becomes complicated, and the curse of dimensionality leads to impractically numerous banks, because the size of a regular lattice grows exponentially with the dimension of space. This will generally be the case in LISA searches of GWs from binaries, since the observed waveforms will be functions of the binary's masses and spin vectors, and of its sky position and orientation.

\paragraph{Stochastic searches.} A possibility, adopted in LIGO's S3 search for spinning binaries \cite{2008PhRvD..78d2002A}, is to employ \emph{stochastic banks} of randomly placed templates \cite{2008arXiv0809.5223M}. Another attractive strategy that avoids banks altogether is to generate one or more stochastic paths that roam about parameter space, visiting different parameter sets with probabilities proportional to the Bayesian posterior probability of a true signal having those parameters. This can be achieved with Markov-Chain Monte Carlo (MCMC) schemes\footnote{The literature on the application of MCMCs to GW detection is too vast to cite here. To my knowledge, the first contribution on this subject is \cite{1998PhRvD..58h2001C}, in the context of enabling Bayesian parameter inference after the detection of binary inspirals.}
\cite{MonteCarlo}:
at every step (\emph{sample}) along the path (\emph{chain}), a proposal is made for the next location to visit by drawing from a distribution that is a function only of the current location; the proposal is then accepted with a probability proportional to the ratio of posteriors for the proposed and current locations. MCMCs are in fact \emph{integration} schemes that yield the moments of the posterior, and therefore the uncertainties in the determination of source parameters. Their convergence varies from run to run, and depends crucially on the choice of proposal distributions, but roughly speaking the schemes converge as one over the square root of the number of \emph{independent} locations; since nearby samples along the path are not independent, the final computations can be made on a subset of samples separated by the \emph{correlation time} of the chain, usually several tens of steps.

Typical MCMC runs for GW parameter-estimation problems visit millions of parameter sets \cite{2006CQGra..23S.819W2006CQGra..23S.761C}; however, if the focus shifts to using MCMCs as searches (that is, to locating the maximum of the posterior), they can be made faster by factors $\sim 100$ by deploying a vast array of tricks \cite{2007CQGra..24.5729C} that accelerate the motion of the chain toward the maximum.\footnote{Most of these tricks do not, as is often stated incorrectly, remove the \emph{Markov property} of the chain (its memorylessness); rather, they destroy the asymptotic convergence of the chain to the target posterior distribution.}
Among these, \emph{annealing} consists in progressively increasing the contrast of the target distribution, starting with a ``hot'' chain that can take bold steps across parameter space, and then cooling it on a preset schedule. In so-called \emph{frequency} annealing, the search starts off with shorter signal templates, which are faster to compute and again reduce contrast; templates are then extended as the search begins to converge. In so-called \emph{thermostated heating}, the dynamic range of the posterior is compressed at its high end, to avoid getting stuck on secondary maxima.  Another useful strategy is reducing the number of parameters by maximizing the posterior analytically over the extrinsic parameters with the ``$\mathcal{F}$'' statistic \cite{1998PhRvD..58f3001J,2004PhRvD..70b2003K}.

As mentioned above, the choice of the proposal distribution is paramount, especially because there can be significant correlations among the changes in the waveforms due to changing different source parameters $\theta^i$. A basic technique (helpful also for parameter estimation) is rescaling local jumps by the eigenvectors and eigenvalues of the Fisher matrix $(\partial h/\partial \theta^i, \partial h/\partial \theta^j)$; strictly speaking, this matrix approximates the curvature of the likelihood surface around a maximum, but here it is used to identify the correct scale of parameter displacements with respect to waveform changes, and to disentangle parameter correlations. The proposals include also long speculative jumps across the entire allowable parameter ranges, and targeted jumps that exploit the nonlocal symmetries of the waveforms (e.g., changing the chirp mass and time of coalescence together to move from the primary posterior maximum to secondary unconnected maxima) or of the LISA response (e.g., moving to antipodal points in the sky, which produce essentially the same LISA response at low frequencies) \cite{2007CQGra..24.5729C}. MCMC schemes have been applied successfully to searching for SMBH binary signals in MLDC 1 \cite{2007CQGra..24..501C2007CQGra..24..521R,2007CQGra..24..595B}, 2 \cite{2008CQGra..25k4037B}, and 1B \cite{2008CQGra..25r4026B}.

The fact that MCMC searches are so efficient in locating the maximum of the posterior indicates that it is not found on a narrow, isolated peak, but rather on a larger mountain that can be climbed semirandomly by the chain. It should be then possible to exploit such a structure to organize a template-bank--based search of efficiency comparable to MCMCs, using hierarchical banks that become increasingly dense and concentrated in promising regions, or that are laid out alternatingly along separate subsets of parameters.\footnote{Conversely, if the maximum was really on an isolated peak, MCMC and bank methods alike would require a number of templates on the order of the ratio between the volume of parameter space and the footprint of the peak.} Indeed, the authors of \cite{2008CQGra..25r4027H} implemented both strategies (as well as stochastic template placement) for their MLDC 1B entry, and succeeded in limiting template evaluations to a few $10^4$, as were needed in MCMC searches; \cite{2008CQGra..25s5011B} achieves a similar result using sophisticated stochastic template placement with location-dependent densities. As mentioned above, bank-based searches have a reassuring value in searching exhaustively over the entire parameter space; it will be interesting to see if they can be extended to spinning-binary signals, which bring in several new parameters. We note also that bank-based searches do benefit from a final MLDC refinement of the best-fit parameters (see also \cite{2007CQGra..24..595B} for an MLDC 1 composite search that joins TF track fitting, a bank-based search, and a final MCMC).


\section{Extreme--mass-ratio inspirals}
\label{sec:emri}

EMRIs involve compact stellar objects in galactic nuclei, which have entered a region of phase space where their further evolution is dominated by GW emission; thus EMRIs are by definition doomed to inspiral and plunge into the central galactic SMBH. The requirement that EMRIs be detectable by LISA constrains the SMBH mass to $10^4\mbox{--}10^7 \, M_\odot$, and the nature of the compact object to a neutron star, white dwarf, or stellar-mass BH, which can approach the SMBH closely without being disrupted. See \cite{2007CQGra..24..113A} for a review of EMRI astrophysics.

EMRIs will typically emit GWs in the LISA band for $\sim 10^5$ cycles, over times comparable to the LISA lifetime, and with signal strengths an order of magnitude below the LISA noise, and a factor of a few deeper below the Galactic-binary foreground. EMRI signals can be understood as the waveforms emitted by test particles moving on geodesics and creating small perturbations ($\propto$ the mass ratio) in the Kerr geometry of the central spinning SMBH. The perturbations travel out to infinity as GWs, but they also scatter off the background geometry and act back on the test particle, which is slightly deviated off its geodesic: this is the ``self-force'' picture of EMRI motion. Indeed, it is BH perturbation theory that enables the computation of EMRI orbits and waveforms: by contrast, EMRI orbital velocities are too relativistic for post-Newtonian calculations to be applicable, whereas the large number of orbits makes numerical-relativity runs unaffordable.

\paragraph{EMRI waveforms.}
See Drasco \cite{2006CQGra..23S.769D} for a taxonomy of the approaches that have been developed to compute EMRI waveforms, from the most accurate (but not yet achieved) self-force waveforms, to Barack and Cutler's ``analytical kludge'' waveforms \cite{2004PhRvD..69h2005B}, which enhance Keplerian orbits (and the corresponding Peters--Mathews GW emission \cite{1963PhRv..131..435P}) by adding post-Newtonian pericenter-- and orbital-plane--precession terms. These waveforms capture the \emph{qualitative} behavior of EMRI waveforms, such as the GW emission at multiple harmonics of the radial frequency modified by the frequency of periastron precession, each with sidebands at multiples of the frequency of orbital-plane precession.
While parameter estimation may eventually require the waveforms on the more accurate end of Drasco's classification, recent investigations of EMRI detection (including those in MLDC 2 and 1B) have used analytic-kludge waveforms as stand-ins, because they can be generated much more rapidly, and they are more transparent to insight on the effect of the source parameters on the evolution of frequencies.

\paragraph{Semicoherent searches.}
The first quantitative investigation of EMRI searches \cite{2004CQGra..21S1595G} concluded that, because of the large number of source parameters (fourteen, of which perhaps seven are especially important), and of the large number of GW cycles that must be matched coherently in a matched-filtering search, it would be unfeasible to deploy a template bank with all the required signal shapes. The authors of \cite{2004CQGra..21S1595G} suggested instead a \emph{semicoherent} search whereby bank-based matched filtering would be applied on shorter data segments of three weeks, and the resulting SNRs summed incoherently along paths in parameter space corresponding to the expected evolution of EMRIs along the inspiral. This strategy is analogous to the \emph{stack--slide} method developed for all-sky searches of periodic sources with LIGO \cite{2000PhRvD..61h2001B}. Template banks would still be very large ($10^{12}$), and the computational load very taxing (many teraflops yr). The reach (maximum detection distance) of such a search would be roughly half the reach of optimal filtering, but would still allow tens to hundreds of detections.

\paragraph{Time--frequency searches.}
More efficient if less powerful searches can be carried out by looking for the tracks left by EMRI signals in the time--frequency (TF) spectrograms of LISA data. Such a program has been developed over the last few years by Wen, Gair, and coauthors, who began by looking for single bright pixels in the TF plane \cite{wengairchen}
(in analogy to the \emph{excess power} method for ground-based searches \cite{2001PhRvD..63d2003A}), using multiple choices of pixel timespan and frequency bandwidth. Such a search has roughly half the reach of the semicoherent strategy, and it is clearly limited by its focus on the brightest single-harmonic, quasimonochromatic stretch of EMRI signal.
This limit is overcome by an extension of the method, the Hierarchical Algorithm for Clusters and Ridges \cite{2007CQGra..24.1145G}, which groups nearby bright pixels in clusters, much like the TFClusters algorithm \cite{2002PhRvD..66j2004S} used in LIGO burst searches, enhancing the reach of the single-pixel method by up to 20\%. In addition, the shape of the clusters carries much useful information, which can be used
\cite{gairmandel}
to estimate the intrinsic parameters of the EMRI (from the evolution of the frequencies and the separation of sidebands) and its sky position (from the modulations in the strength of the track in the different TDI observables throughout the inspiral). The extended method was applied to the EMRI Challenges in MLDC 2 and 1B
\cite{gairmandel},
where it was able to estimate all intrinsic parameters (and for MLDC 1B, also the sky position) to a few percent, and much better for the fundamental radial frequency.

Altogether, TF searches are very fast (they run in minutes); they are applicable also to other bright LISA sources, including SMBH and Galactic binaries \cite{2007CQGra..24.1145G}; and they may serve as a first stage in a hierarchical search, segueing into targeted matched filtering. However, all explorations so far have been for isolated sources in Gaussian LISA noise, and there is a question of how performance may be reduced in the realistic case, with multiple intersecting tracks from instrument glitches and other strong sources (SMBH binary inspirals, the brightest Galactic binaries, and multiple simultaneous EMRIs). Even if the track-identification problem proves intractable in general, the brightest sources will still stand out above the confusion.

\paragraph{Stochastic searches.} The high dimension of EMRI parameter space calls naturally for a stochastic approach to matched filtering. Indeed, MCMC codes developed to search for SMBH binaries in MLDC data have been adapted for EMRI waveforms \cite{2008CQGra..25r4030G,2008arXiv0804.3323C} (see also the earlier example \cite{2006AIPC..873..444S}). The main problem in the adaption is to secure a sufficiently broad sampling of parameter space, so that the chain is given a chance to wander sufficiently close to the true posterior maximum, and at the same to avoid getting stuck on a secondary maximum---there are as many of these as there are ways for the multiple TF tracks of two EMRI signals to find themselves in partial constructive interference for widely separated parameter sets.

To this purpose, the authors of \cite{2008CQGra..25r4030G,2008arXiv0804.3323C} have again employed tricks ranging from temperature annealing to time annealing (equivalent to frequency annealing for SMBHs) and thermostated heating. Multiple proposal distributions are used, including long speculative jumps followed by a reconnaissance search in the neighborhood of the landing site before deciding whether to accept the jump \cite{2008arXiv0804.3323C}; jumps limited to intrinsic, extrinsic, or phase parameters \cite{2008CQGra..25r4030G}; and ``island hopping,'' constrained jumps designed to maintain gross waveform features such as track geometry. The implementation of this feature in \cite{2008CQGra..25r4030G} is especially interesting: strong harmonic features are first identified with small MCMC runs; the subsequent search is constrained to parameter sets that would reproduce them, using parameter relations valid for analytic-kludge waveforms.

Trials on MLDC datasets \cite{2008CQGra..25r4030G,2008arXiv0804.3323C,2008CQGra..25r4026B} have shown that MCMC searches are capable of detecting moderately strong, isolated signals with good parameter-estimation performance. Much as for TF searches, it remains to be seen which of the strategies discussed here will be effective
for weaker EMRI signals, and in the presence of confusion noise, of other LISA sources, and of other EMRIs. (MLDC 3 does include a search for five simultaneous, weaker EMRIs.) As in the case of SMBH searches, convergence-acceleration techniques are incompatible with the estimation of the full posterior, so subsequent, longer runs without tricks will be needed. If the mountains around the secondary maxima hold significant probability, they must be included in parameter uncertainties in addition to the width of the global maximum, so we will need to ensure that the parameter-estimation MCMCs visit them fairly. It is likely that some sort of island hopping will be needed for this purpose; thus, it is important to develop an understanding of the (approximate) discrete symmetries of EMRI parameter space that conserve the properties of EMRI signals. See \cite{2007arXiv0711.4644D} for the beginnings of such an understanding.

\section{Galactic binaries}
\label{sec:gbs}

The dominant GW contribution in the LISA band will come from short-period binaries of compact stellar objects (white dwarfs, neutron stars, black holes, and naked He stars), predominantly from the tens of millions of such systems in our Galaxy \cite{1997CQGra..14.1439B,2001A&A...375..890N}, but also from farther afield
\cite{extragalactic}.
We know from astronomical observations that a few known Galactic binaries are sufficiently strong to be detected by LISA with the matched-filtering SNR accreted in a few months after commissioning \cite{2006CQGra..23S.809S}; these \emph{verification binaries} will provide the first test of the correct operation of the instrument.

Throughout its lifespan, LISA will be able to detect and characterize a few $10^4$ additional unknown \emph{resolvable binaries} \cite{2006PhRvD..73l2001T}, but most systems will be either too weak or too crowded to be detected individually, and will form a \emph{confusion-limited foreground} that will act as effective noise for the detection of other sources. Confusion will begin roughly below $\sim 3$ mHz, where the binary density reaches 0.5 per FFT bin \cite{1997CQGra..14.1439B,2004PhRvD..70l2002B}. Thus, most resolved binaries will be found above 3 mHz, with some at lower frequencies if they are sufficiently massive or close to us that they stand out from the crowd.  After removing these, the confusion foreground is left with statistics very similar to those of colored Gaussian noise \cite{2006PhRvD..73l2001T}, with some imprints of having been emitted from the direction of the Galaxy \cite{2005PhRvD..71l2003E}.

\paragraph{Matched filtering.}
Compared to SMBHs and EMRIs, Galactic-binary signals are simple: by the time they enter the LISA band, orbits have overwhelmingly become circularized, and frequency evolution is slow enough that it can be matched accurately by a linear model. However, the resulting quasimonochromatic signals are modulated by the LISA response: the LISA motion around the Sun smears the central frequency $f$ into a bandwidth of $300 \times (f/3\, \mathrm{mHz})$ nHz, and the rotation of the LISA triangle creates eight sidebands, displaced symmetrically by frequency multiples of 1/year $\sim 32$ nHz (see, e.g., \cite{2004PhRvD..70b2003K}). Thus, simple spectral analysis is insufficient, and matched filtering again comes to the fore.  The $\mathcal{F}$ statistic can be used to maximize the likelihood analytically over the extrinsic parameters, so that templates can be distributed only over the remaining three ($f$ and the sky-position angles), plus $\dot{f}$ if $f \gtrsim 1.6$ mHz. Indeed, Prix, Whelan, and Khurana ran such searches on MLDC 1, 2, and 1B data \cite{2007CQGra..24..565P2008CQGra..25r4029W,mldcreports}, adapting software from the LIGO searches for continuous GWs, and implementing a sophisticated pipeline that includes a first wide-parameter $\mathcal{F}$-statistic search run separately on multiple TDI channels, a coincidence sieve, and a final zoomed-in search that is run in parallel over the TDI channels.

\paragraph{Source confusion.} Unfortunately, any search that treats binary signals individually must contend with the effects of source confusion: signals that are close in parameter space have nonzero correlations, so the template bank will not ``ring'' at the correct parameter locations, but rather at points displaced by systematic confusion errors. A local analysis of the multisource likelihood surface shows that parameter-estimation accuracy decays exponentially with the number of overlapping signals, and superexponentially with the degree of correlation \cite{2004PhRvD..70h2004C}. A strong source may also hide a weaker one, or two sources of comparable strength may fuse into one.

Iterative detection schemes have been developed to deal with this problem: gCLEAN identifies the brightest source in the data and subtracts a small fraction of it; this is repeated until a prescribed residual power is reached, and all signals are then reconstructed from the record of subtractions \cite{2003PhRvD..67j3001C}. ``Slice and Dice'' again identifies the brightest source, but subtracts it completely; then the parameters of all the sources identified so far are refit simultaneously by a global least squares, and resubtracted with the corrected parameters to provide the data for the next step; the process ends when the residual becomes consistent with noise \cite{2006AIPC..873..489R}.

\paragraph{Global fits.} A more radical approach is searching for all sources at the same time, by maximizing the posterior probability over a $(N \times M)$-dimensional manifold, where $N$ is the number of sources and $M$ the number of parameters. Template-bank schemes to do so become quickly unmanageable, so once again the community turned to MCMCs. Umstaetter and colleagues \cite{2005PhRvD..72b2001U} tackled an idealized problem that does not include the LISA response; their paper is a very good introduction to the statistical underpinnings of such a search. Cornish and Crowder were then able to develop an effective MCMC LISA search, at first for a few binaries \cite{2005PhRvD..72d3005C}, then for a realistic number of resolvable systems \cite{2007PhRvD..75d3008C}. The crucial realization is that signals that are well separated in frequency have no significant overlap, so the global fit of $\sim 10^4$ binaries can be tackled in frequency-domain blocks that contain $\sim 100$ sources, taking special precautions not to miss those at the boundaries between blocks. Once again, the convergence of the chains can be accelerated with tricks such as annealing, multiple proposal distributions, and source-tailored special moves. The Cornish--Crowder code was spectacularly demonstrated in MLDC 2, where it identified 19,324 binaries out of $\sim 26 \times 10^6$; template-bank searches were able identify roughly a factor of ten fewer \cite{2008CQGra..25k4037B}.

\paragraph{Heterodox alternatives.}
In closing, I like to mention \emph{genetic searches} (\cite{2006PhRvD..73f3011C}, demonstrated in MLDC 1 \cite{2007CQGra..24..575C}), which are similar in spirit to MCMC, but espouse metaphors of mutation and selection rather than those of statistical physics; and \emph{tomographic recostruction} (\cite{2006PhRvD..73h3006M2006AIPC..873..429N}, applied to MLDC 1 and 2 \cite{2007CQGra..24..587N,2008CQGra..25k4037B}), which exploits the fact that the instantaneous LISA response can be expressed as a \emph{Radon transform} \cite{deans1983} of the signal from all Galactic binaries (roughly speaking, an integral over all sky locations at times retarded by propagation to the LISA position); the transform can be inverted, yielding a somewhat smeared map of the sky as a function of frequency.

\section{In conclusion}

The heritage from ground-based detectors and the demonstrations given in the MLDCs have already gone a long way in providing solid foundations for the LISA science objectives. LISA data analysis is a vibrantly active field with many intellectual challenges still open, and the MLDCs have proved a stimulating platform to develop and test new and established ideas alike. MLDC contributions (covered only partially in this article) provide an ideal springboard to begin exploring the field.

\ack
\vspace{-0.2cm}
MV was supported by the LISA Mission Science Office and by JPL's HRDF and RTD programs. This work was carried out at the Jet Propulsion Laboratory, California Institute of Technology, under contract with the National Aeronautics and Space Administration.

\section*{References}

\newcommand{\prd}{Phys. Rev. D}
\newcommand{\apj}{Astrophys. J.}
\newcommand{\apjs}{Astrophys. J. Suppl. Ser.}
\newcommand{\apjl}{Astrophys. J. Lett.}
\newcommand{\mnras}{Mon. Not. R. Astron. Soc.}
\newcommand{\aap}{A\&A}
\newcommand{\jcp}{J. Chem. Phys.}


\begin{thebibliography}{10}
\expandafter\ifx\csname url\endcsname\relax
  \def\url#1{{\tt #1}}\fi
\expandafter\ifx\csname urlprefix\endcsname\relax\def\urlprefix{URL }\fi
\providecommand{\eprint}[2][]{\url{#2}}

\bibitem{lisa}
Bender P, Danzmann P and the LISA Study~Team 1998 ``Laser Interferometer Space
  Antenna for the detection of gravitational waves, pre-phase A report'' MPQ \textbf{233} (Garching: Max-Planck-Instit\"ut f\"ur Quantenoptik)

\bibitem{lisascience}
{LISA Mission Science Office} 2007 ``LISA: Probing the universe with
  gravitational waves'' \url{www.lisascience.org/resources/talks-articles/science/%
lisa_science_case.pdf}

\bibitem{lisa6} {Merkovitz} S~M and {Livas} J~C (eds) 2006 {\em Laser Interferometer Space Antenna: 6th International LISA Symposium\/} ({\em American Institute of Physics Conference Series\/} vol 873)

\bibitem{mldc1form} {Arnaud} K~A et al. 2006 in \cite{lisa6} p 619, 625

\bibitem{2007CQGra..24..529A} {Arnaud} K~A et al. 2007 {\em Classical and Quantum Gravity\/} {\bf 24} 529

\bibitem{2007CQGra..24..551A}
{Arnaud} K~A et al. 2007 {\em Classical and Quantum Gravity\/} {\bf 24} 551

\bibitem{2008CQGra..25k4037B}
{Babak} S et al. 2008 {\em Classical and Quantum Gravity\/} {\bf
  25} 114037

\bibitem{2008CQGra..25r4026B}
{Babak} S et al. 2008 {\em Classical and Quantum Gravity\/} {\bf 25} 184026

\bibitem{mldcreports}
MLDC entries, reports, and challenge keys \url{www.tapir.caltech.edu/~mldc}

\bibitem{keycornish2008} Key Shapiro J and Cornish N J 2008 arXiv.org:0812.1590

\bibitem{stochasticprinciple}
{Tinto} M, {Armstrong} J~W and {Estabrook} F~B 2001 {\em \prd\/} {\bf 63} 021101;
{Hogan} C~J and {Bender} P~L 2001 {\em \prd\/} {\bf 64} 062002;
{Sylvestre} J and {Tinto} M 2003 {\em \prd\/} {\bf 68} 102002

\bibitem{2008CQGra..25r4019R}
{Robinson} E~L, {Romano} J~D and {Vecchio} A 2008 {\em Classical and Quantum Gravity\/} {\bf 25} 184019

\bibitem{stochasticanisotropy}
{Giampieri} G and {Polnarev} A~G 1997 {\em Classical and Quantum Gravity\/} {\bf 14} 1521;
{Ungarelli} C and {Vecchio} A 2001 {\em \prd\/} {\bf 64} 121501;
{Seto} N and {Cooray} A 2004 {\em \prd\/} {\bf 70} 123005;
{Kudoh} H and {Taruya} A 2005 {\em \prd\/} {\bf 71} 024025;
{Taruya} A and {Kudoh} H 2005 {\em \prd\/} {\bf 72} 104015;
{Taruya} A 2006 {\em \prd\/} {\bf 74} 104022;
{Seto} N 2006 {\em Physical Review Letters\/} {\bf 97} 151101

\bibitem{1999PhT....52j..44B}
{Barish} B~C and {Weiss} R 1999 {\em Physics Today\/} {\bf 52} 44

\bibitem{2005PhRvD..72d2003V}
{Vallisneri} M 2005 {\em \prd\/} {\bf 72} 042003

\bibitem{2006ApJS..163...50H}
{Hopkins} P~F et al. 2006
  {\em \apjs\/} {\bf 163} 50

\bibitem{2007MNRAS.377.1711S}
{Sesana} A, {Volonteri} M and {Haardt} F 2007 {\em \mnras\/} {\bf 377}
  1711

\bibitem{ligosearch} {Abbott} B et al. 2006 {\em \prd\/} {\bf 73} 062001; {Abbott} B et al. 2008 {\em \prd\/} {\bf 77} 062002

\bibitem{2008PhRvD..78d2002A} {Abbott} B et al. 2008 {\em \prd\/} {\bf 78} 042002

\bibitem{breakthrough} {Pretorius} F 2005 {\em Physical Review Letters\/} {\bf 95} 121101; {Campanelli} M et al. 2006 {\em Physical Review Letters\/} {\bf 96} 111101; {Baker} J~G et al. 2006 {\em Physical Review Letters\/} {\bf 96} 111102

\bibitem{pretoriusreview}
{Pretorius} F 2009 in {\em Physics of Relativistic Objects in Compact Binaries: from Birth to Coalescence\/} ({\em Astrophysics and Space Science Library\/} vol 359) eds Colpi M et al. (Springer-Verlag)

\bibitem{stitched} {Buonanno} A et al. 2007 {\em \prd\/} {\bf 76} 104049; {Ajith} P et al. 2007 {\em Classical and Quantum Gravity\/} {\bf 24} 689; {Ajith} P et al. 2008 {\em \prd\/} {\bf 77} 104017; {Buonanno} A 2008 in {\em Astrophysics of Compact Objects\/} ({\em American Institute of Physics Conference Series\/} vol 968) ed {Yuan} Y~F, {Li} X~D and {Lai} D p 307

\bibitem{ninjasite} NINJA project homepage \url{www.gravity.phy.syr.edu/dokuwiki/doku.php?id=ninja:hom%
e}

\bibitem{2006CQGra..23S.785B}
{Berti} E 2006 {\em Classical and Quantum Gravity\/} {\bf 23} 785

\bibitem{2007PhRvD..75l4002A}
{Arun} K~G et al. 2007 {\em \prd\/}
  {\bf 75} 124002

\bibitem{2007PhRvD..76j4018C}
{Cutler} C and {Vallisneri} M 2007 {\em \prd\/} {\bf 76} 104018

\bibitem{basics} {Finn} L~S 1992 {\em \prd\/} {\bf 46} 5236; {Finn} L~S and {Chernoff} D~F 1993 {\em \prd\/} {\bf 47} 2198; {Cutler} C and {Flanagan} {\'E}~{\'E} 1994 {\em \prd\/} {\bf 49} 2658

\bibitem{Bender:2006rz}
{Bender} P~L et al. 2006 ``LISA data issues task force report'' \url{www.srl.caltech.edu/lisa/mission_documents.html}

\bibitem{2008arXiv0809.5223M}
{Messenger} C, {Prix} R and {Papa} M~A 2008 arXiv.org:0809.5223

\bibitem{1998PhRvD..58h2001C}
{Christensen} N and {Meyer} R 1998 {\em \prd\/} {\bf 58} 082001

\bibitem{MonteCarlo} {Metropolis} N et al. 1953 {\em \jcp\/} {\bf 21} 1087;
Liu J~S 2001 {\em Monte Carlo strategies in scientific computing\/} (New York: Springer)

\bibitem{2006CQGra..23S.819W2006CQGra..23S.761C} {Wickham} E~D~L, {Stroeer} A and {Vecchio} A 2006 {\em Classical and Quantum Gravity\/} {\bf 23} 819; {Cornish} N~J and {Porter} E~K 2006 {\em Classical and Quantum Gravity\/} {\bf 23} 761

\bibitem{2007CQGra..24.5729C}
{Cornish} N~J and {Porter} E~K 2007 {\em Classical and Quantum Gravity\/} {\bf
  24} 5729

\bibitem{1998PhRvD..58f3001J}
{Jaranowski} P, {Kr{\'o}lak} A and {Schutz} B~F 1998 {\em \prd\/} {\bf 58}
  063001

\bibitem{2004PhRvD..70b2003K}
{Kr{\'o}lak} A, {Tinto} M and {Vallisneri} M 2004 {\em \prd\/} {\bf 70}
  022003

\bibitem{2007CQGra..24..501C2007CQGra..24..521R} {Cornish} N~J and {Porter} E~K 2007 {\em Classical and Quantum Gravity\/} {\bf 24} 501; {R{\"o}ver} C et al. 2007 {\em Classical and Quantum Gravity\/} {\bf 24} 521

\bibitem{2007CQGra..24..595B} {Brown} D~A et al. 2007 {\em Classical and Quantum Gravity\/} {\bf 24} 595

\bibitem{2008CQGra..25r4027H}
{Harry} I~W, {Fairhurst} S and {Sathyaprakash} B~S 2008 {\em Classical and
  Quantum Gravity\/} {\bf 25} 184027

\bibitem{2008CQGra..25s5011B}
{Babak} S 2008 {\em Classical and Quantum Gravity\/} {\bf 25} 195011

\bibitem{2007CQGra..24..113A}
{Amaro-Seoane} P et al. 2007 {\em Classical and Quantum Gravity\/} {\bf 24} 113

\bibitem{2006CQGra..23S.769D}
{Drasco} S 2006 {\em Classical and Quantum Gravity\/} {\bf 23} 769

\bibitem{2004PhRvD..69h2005B}
{Barack} L and {Cutler} C 2004 {\em \prd\/} {\bf 69} 082005

\bibitem{1963PhRv..131..435P}
{Peters} P~C and {Mathews} J 1963 {\em Physical Review\/} {\bf 131} 435

\bibitem{2004CQGra..21S1595G}
{Gair} J~R et al. 2004 {\em Classical and Quantum Gravity\/} {\bf 21}
  1595

\bibitem{2000PhRvD..61h2001B}
{Brady} P~R and {Creighton} T 2000 {\em \prd\/} {\bf 61} 082001

\bibitem{wengairchen} {Wen} L and {Gair} J~R 2005 {\em Classical and Quantum Gravity\/} {\bf 22} 445; {Gair} J and {Wen} L 2005 {\em Classical and Quantum Gravity\/} {\bf 22} 1359; {Wen} L, {Chen} Y and {Gair} J 2006 in \cite{lisa6} p 595


\bibitem{2001PhRvD..63d2003A}
{Anderson} W~G et al. 2001 {\em \prd\/} {\bf 63} 042003

\bibitem{2007CQGra..24.1145G}
{Gair} J and {Jones} G 2007 {\em Classical and Quantum Gravity\/} {\bf 24}
  1145

\bibitem{2002PhRvD..66j2004S}
{Sylvestre} J 2002 {\em \prd\/} {\bf 66} 102004

\bibitem{gairmandel} {Gair} J~R, {Mandel} I and {Wen} L 2008 {\em Journal of Physics Conference Series\/} {\bf 122} 012037; {Gair} J~R, {Mandel} I and {Wen} L 2008 {\em Classical and Quantum Gravity\/} {\bf 25} 184031


\bibitem{2008CQGra..25r4030G} {Gair} J~R, {Porter} E, {Babak} S and {Barack} L 2008 {\em Classical and Quantum Gravity\/} {\bf 25} 184030

\bibitem{2008arXiv0804.3323C} {Cornish} N~J 2008 arXiv.org:0804.3323

\bibitem{2006AIPC..873..444S}
{Stroeer} A, {Gair} J and {Vecchio} A 2006 in \cite{lisa6} p 444

\bibitem{2007arXiv0711.4644D}
{Drasco} S 2007 arXiv.org:0711.4644

\bibitem{1997CQGra..14.1439B} {Bender} P~L and {Hils} D 1997 {\em Classical and Quantum Gravity\/} {\bf 14} 1439

\bibitem{2001A&A...375..890N} {Nelemans} G, {Yungelson} L~R and {Portegies Zwart} S~F 2001 {\em \aap\/} {\bf 375} 890

\bibitem{extragalactic} {Benacquista} M~J 2006 {\em Living Reviews in Relativity\/} {\bf 9} 2; {Cooray} A and {Seto} N 2005 {\em \apjl\/} {\bf 623} L113; {Farmer} A~J and {Phinney} E~S 2003 {\em \mnras\/} {\bf 346} 1197

\bibitem{2006CQGra..23S.809S}
{Stroeer} A and {Vecchio} A 2006 {\em Classical and Quantum Gravity\/} {\bf 23}
  809

\bibitem{2006PhRvD..73l2001T}
{Timpano} S~E, {Rubbo} L~J and {Cornish} N~J 2006 {\em \prd\/} {\bf 73}
  122001

\bibitem{2004PhRvD..70l2002B}
{Barack} L and {Cutler} C 2004 {\em \prd\/} {\bf 70} 122002

\bibitem{2005PhRvD..71l2003E}
{Edlund} J~A et al. 2005 {\em \prd\/} {\bf
  71} 122003

\bibitem{2007CQGra..24..565P2008CQGra..25r4029W} {Prix} R and {Whelan} J~T 2007 {\em Classical and Quantum Gravity\/} {\bf 24} 565; {Whelan} J~T, {Prix} R and {Khurana} D 2008 {\em Classical and Quantum Gravity\/} {\bf 25} 184029

\bibitem{2004PhRvD..70h2004C}
{Crowder} J and {Cornish} N~J 2004 {\em \prd\/} {\bf 70} 082004

\bibitem{2003PhRvD..67j3001C}
{Cornish} N~J and {Larson} S~L 2003 {\em \prd\/} {\bf 67} 103001

\bibitem{2006AIPC..873..489R}
{Rubbo} L~J, {Cornish} N~J and {Hellings} R~W 2006 in in \cite{lisa6} p 489

\bibitem{2005PhRvD..72b2001U}
{Umst{\"a}tter} R et al. 2005 {\em \prd\/} {\bf 72} 022001

\bibitem{2005PhRvD..72d3005C}
{Cornish} N~J and {Crowder} J 2005 {\em \prd\/} {\bf 72} 043005

\bibitem{2007PhRvD..75d3008C}
{Crowder} J and {Cornish} N~J 2007 {\em \prd\/} {\bf 75} 043008

\bibitem{2006PhRvD..73f3011C}
{Crowder} J, {Cornish} N~J and {Reddinger} J~L 2006 {\em \prd\/} {\bf 73}
  063011

\bibitem{2007CQGra..24..575C}
{Crowder} J and {Cornish} N~J 2007 {\em Classical and Quantum Gravity\/} {\bf
  24} 575

\bibitem{2006PhRvD..73h3006M2006AIPC..873..429N} {Mohanty} S~D and {Nayak} R~K 2006 {\em \prd\/} {\bf 73} 083006; {Nayak} K~R, {Mohanty} S~D and {Hayama} K 2006 in \cite{lisa6} p 429


\bibitem{2007CQGra..24..587N}
{Nayak} K~R, {Mohanty} S~D and {Hayama} K 2007 {\em Classical and Quantum
  Gravity\/} {\bf 24} 587

\bibitem{deans1983}
Deans S~R 1983 {\em The Radon transform and some of its applications\/} (New
  York: Wiley)

\end{thebibliography}

\providecommand{\newblock}{}

\end{document}